\documentclass[sn-mathphys,Numbered]{sn-jnl}

\usepackage{graphicx}%
\usepackage{multirow}%
\usepackage{amsmath,amssymb,amsfonts}%
\usepackage{amsthm}%
\usepackage{mathrsfs}%
\usepackage[title]{appendix}%
\usepackage{xcolor}%
\usepackage{textcomp}%
\usepackage{manyfoot}%
\usepackage{booktabs}%
\usepackage{algorithm}%
\usepackage{algorithmicx}%
\usepackage{algpseudocode}%
\usepackage{listings}%

\theoremstyle{thmstyleone}

\theoremstyle{thmstyletwo}

\theoremstyle{thmstylethree}

\begin{document}

\title[Article Title]{HyMo: Vulnerability Detection in Smart Contracts using a Novel Multi-Modal Hybrid Model}

\author{\fnm{Mohammad} \sur{Khodadadi}}

\author*[1]{\fnm{Jafar} \sur{Tahmoresnezhad}}\email{j.tahmores@it.uut.ac.ir}

\affil[1]{\orgdiv{Department of IT \& Computer Engineering}, \orgname{Urmia University of Technology}, \orgaddress{\city{Orūmīyeh}, \country{Iran}}}

\abstract{With blockchain technology rapidly progress, the smart contracts have become a common tool in a number of industries including finance, healthcare, insurance and gaming. The number of smart contracts has multiplied, and at the same time, the security of smart contracts has drawn considerable attention due to the monetary losses brought on by smart contract vulnerabilities. Existing analysis techniques are capable of identifying a large number of smart contract security flaws, but they rely too much on rigid criteria established by specialists, where the detection process takes much longer as the complexity of the smart contract rises. In this paper, we propose HyMo as a multi-modal hybrid deep learning model, which intelligently considers various input representations to consider multimodality and FastText word embedding technique, which represents each word as an n-gram of characters with BiGRU deep learning technique, as a sequence processing model that consists of two GRUs to achieve higher accuracy in smart contract vulnerability detection. The model gathers features using various deep learning models to identify the smart contract vulnerabilities. Through a series of studies on the currently publicly accessible dataset such as ScrawlD, we show that our hybrid HyMo model has excellent smart contract vulnerability detection performance. Therefore, HyMo performs better detection of smart contract vulnerabilities against other approaches.}


\maketitle

\section{Introduction}\label{introduction}

The smart contracts were introduced in 1994 for the first time \citep{mohanta2018overview}. The smart contracts are a set of digital agreements that contain information about how to participate in fulfilling those agreements. The purpose of smart contract is to implement the content of contracts through cryptographic protocols and digital security mechanisms. However, due to the technological limitations, it was not possible to deploy a smart contract until someone named Satoshi Nakamoto introduced the Bitcoin, which works on the basis of blockchain technology where the smart contracts can be deployed on it \citep{kawaguchi2019application}.

In essence, a blockchain network's miners maintain a distributed and shared transaction ledger via a consensus protocol. All transactions must be immutable after being recorded on the blockchain thanks to the consensus mechanism and duplicated ledgers. A smart contract is a self-executing, self-verifying contract. Although it contains intricate time and order dependencies, flaws in a smart contract's logic and syntax makes it vulnerable, which leads to improper automated execution \citep{zheng2017blockchain, zou2019smart}.

The Ethereum virtual machine (EVM) has made it possible to execute distributed programs in form of smart contracts. In fact, EVM is a completely stack-based virtual machine that supports 134 opcodes in its instruction set and can run Turing-complete programs. Typically, smart contract code is written by developers in a high-level language that are converted to EVM bytecode \citep{solidity}.

Arithmetic flaws, such as integer overflow and underflow, are a common type of weakness in many programming languages, but they can have very serious repercussions in the context of smart contracts. For instance, if a loop counter overflows and generates an infinite loop, the funds of the smart contract may be completely frozen. The loop can be exploited by the  attacker if they know how to increase the number of iterations, for adding the additional users to the smart contract.

Ethereum is one of the most widely used blockchain platforms \citep{liao2019soliaudit}. An Ethereum account known as an external account, which manages the amount of Ethereum currency with a value in the billions of dollars, is what makes up the smart contract itself. Amounts of money of this kind may potentially draw the attention of potential attackers. In 2016, a flaw in the DAO's smart contract led to the loss of Ether valued at USD 55 million \citep{liu2019survey}. In 2017, a smart contract flaw in the parity wallet resulted in the loss of more than USD 30 million worth of Ether \citep{qian2020towards}. Such security concerns pose the significant challenges for the growth of smart contracts and result in a crisis of trust between users and smart contracts. Therefore, it is imperative to have a reliable method for identifying smart contract vulnerabilities.

In this research, we suggest HyMo as a multi-modal deep learning approach to identify the smart contract vulnerabilities by combining two input representations (i.e., cleaned source code and compiled source code) to consider multimodality, FastText word embedding model, to represent each word as an n-gram of characters and BiGRU deep learning model, as a sequence processing model that consists of two GRUs where one taking the input in a forward direction, and the other in a backwards direction. Our solution is usable for all types of smart contract vulnerabilities but in this paper, we focus on arithmetic vulnerabilities.

In general, the main contributions of our work are as follows.

\begin{enumerate}
\item Our HyMo model increase the accuracy of smart contract vulnerability detection by using the various input representations, deep learning techniques and word embedding techniques.

\item The smart contract has two representations as an input, one cleaned source code and the other is compiled source code as opcode.

\item Our tests and results are more realistic because our dataset incorporates real-world smart contracts.

\item The tests are carried out to use the benefits of different input representations, deep learning and multi-modal hybrid learning to improve the performance of vulnerability identification compared to a single neural network model or hybrid models without the use of different representations.

\item Our model is extensible for all types of smart contract vulnerabilities.

\item We experiment four different hybrid model with different structure on a dataset, which is the combination of ScrawlD \citep{yashavant2022scrawld} and SmartBugs Dataset-Wild \citep{durieux2020empirical} samples and we achieve 79.71\% accuracy given that our dataset contains real-world smart contracts.
\end{enumerate}

In the following, we talk about the previous studies on smart contract vulnerabilities, arithmetic vulnerabilities, how our model works and the achieved results by our model.

\section{Related Work}\label{related_work}

In this section, we review the existing works from two aspects, one is the traditional methods and the other one is the deep learning-based hybrid models.

\subsection{Smart Contract Vulnerability Detection}\label{smart_contract_vulnerability_detection}

Variety of tools for detection of the smart contract vulnerabilities have been developed by researchers. Oyente \citep{luu2016making} is a smart contract vulnerability detection tool based on symbolic execution that proposed by Luu L. et al. Oyente is a symbolic execution tool that works directly with EVM bytecode without access to the high-level representation and identifies seven different types of smart contract vulnerabilities.  Osiris \citep{torres2018osiris} is a framework that uses both symbolic execution and taint analysis methods. Tikhomirov S. et al. \citep{tikhomirov2018smartcheck} suggested the extensible static analysis tool known as SmartCheck. SmartCheck examines the Solidity source code against XPath patterns after converting it to an XML-based intermediate representation. The official smart contract vulnerability detection tool for Ethereum is called Mythril \citep{consensys}. Mythril can identify a wide range of smart contract security flaws and uses the symbolic execution to investigate all potentially dangerous paths. A smart contract static analysis tool called Securify \citep{tsankov2018securify} is used to find the security features in smart contract EVM bytecodes. Slither \citep{feist2019slither} is a static analysis tool that is used to find the smart contract vulnerabilities with taking the Solidity abstract syntax tree (AST) generated by the Solidity compiler as input and generates control flow graph (CFG) to detect vulnerabilities. A fuzzy testing tool called ContractFuzzer \citep{jiang2018contractfuzzer}  proposed for smart contract vulnerabilities that can produce fuzzy test inputs based on API specification of smart contract, where it records the runtime state of the smart contract using an EVM, examines logs, and reports security flaws. The fuzzy testing tools for smart contract vulnerabilities are Echidna \citep{grieco2020echidna} and EthRacer \citep{kolluri2019exploiting}. These techniques primarily use the formal verification, symbolic execution, static analysis, taint analysis, and fuzzy testing. Moreover, they perform the vulnerability detection that rely on rigid logical rules, which established in advance by experts.

Deep learning-based methods recently have been applied in a number of industries including vulnerability detection \citep{xing2020new,zhang2022cbgru,gao2020checking}. Convolutional neural networks (CNNs) were utilized by Harer J.A. et al. \citep{harer2018automated} to classifying the vulnerabilities, learninng the features using neural networks, and extracting the control flow graphs (CFGs) of functions at the function level. Through experiments, Sicong Cao et al. \citep{cao2021bgnn4vd} demonstrated the excellent precision and accuracy of the BGNN4VD model, which performs vulnerability identification by building a bipartite graph neural network. A deep learning-based vulnerability detection system called VulDeePecker \citep{li2018vuldeepecker} proposed by Zhen Li et al. VulDeePecker is used the bi-directional long-short term memory (Bi-LSTM) neural networks to find the vulnerabilities. Deep learning and the detection of smart contract vulnerabilities have also been integrated by many researches \citep{xing2020new,zhang2022cbgru,gao2020checking}. With the concept of vulnerability candidate slicing (VCS) \citep{yu2021deescvhunter}, which has rich semantic and syntactic features that can significantly boost the performance of deep learning smart contract models in smart contract vulnerability detection, Yu X et al. \citep{yu2021deescvhunter} developed a systematic and modular framework for smart contract vulnerability called DeeSCVHunter. In order to capture the key characteristics of contracts before performing the smart contract vulnerability detection, Wu H et al. \citep{wu2021peculiar} used a smart contract representation technique based on important data flow graph information. They also proposed a new tool called Peculiar. By utilizing the crucial data flow graph technique, Peculiar enhances the detection performance, but the process of the created critical data flow graphs is difficult, and Peculiar can only identify the smart contract reentry vulnerabilities. Through extensive trials, Qian P. et al. \citep{qian2020towards} showed that their suggested system, known as BLTM-ATT, can detect the smart contract vulnerabilities more accurately than other approaches by combining a bidirectional long-short term memory network with an attention mechanism. However, the relationship between the word embedding technique and the deep learning model is not taken into account by the BLSTM-ATT model. A new vulnerability detection slicing matrix was proposed by Xing C et al. \citep{xing2020new}, where is empirically shown the increase of accuracy in vulnerability identification. In fact, it segments the contracts' opcodes, to use the "Return" as the segmentation point, where it cannot entirely discriminate between valuable and useless operands, and results in partial feature loss and impairs the model's effectiveness in identifying the smart contract vulnerabilities. A method for autonomously learning the smart contract features based on the character embedding and vector space comparison proposed by Zhipeng Gao et al. \citep{gao2020checking} to find the potential smart contract vulnerabilities, where the fundamental strategy is to parse the smart contract code into a stream of characters with code structure information, turn the code elements into vectors, and assess how closely the coded vectors resemble known problems. TokenCheck \citep{goswami2021tokencheck}, a smart contract vulnerability discovery tool proposed by Goswami S et al., to successfully test on the Ethernet smart contract dataset ERC-20 \citep{ethereum}. For feature extraction, TokenCheck employs a single LSTM neural network.

\subsection{Hybrid Models}\label{hybrid_models}

The hybrid model's function is to integrate each network's advantages in order to get superior outcomes. Hybrid models have been successfully applied in numerous fields. For the purpose of predicting the air quality, Du S et al. \citep{du2019deep} combined one-dimensional convolutional neural networks (1D-CNNs) and bi-directional long-short memory networks (Bi-LSTMs). Convolutional neural networks are used to extract features, and bi-directional long-short memory neural networks are used to learn the correlation of features. In an experiment, Yue W et al. \citep{yue2020sentiment} showed that the hybrid network model outperforms the single-structured neural network in classifying the short texts by combining the word vector model (Word2Vec), the bi-directional long-short term memory network (Bi-LSTM), and the convolutional neural network. In order to use the multilayer convolutional neural network (MLCNN) and the bi-directional gated recurrent unit (BiGRU) with the attention mechanism to news categorization, Duan J et al. \citep{duan2020news} devised a hybrid neural network model (MLCN and BiGRU-ATT). CBGRU model proposed by Zhang L et al. \citep{zhang2022cbgru} to perform the feature extraction. The pre-processing of CBGRU considers the smart contract's integrity to guarantee the semantic consistency of the smart contract.

Our proposed HyMo model uses the appropriate arrangement of various input representations (e.g., cleaned source code and compiled source code), deep learning models (e.g., BiGRU) and word embedding models (e.g., FastText), so it can extract more features and have a better sight on the contents of the smart contract and better accuracy than state-of-the-art of hybrid models in smart contract vulnerability detection.

\section{Raise the Problem}\label{raise_the_problem}

In this section, we discuss the upcoming challenges and vulnerabilities of the smart contracts, especially arithmetic vulnerabilities, and then express our motivation for proposing HyMo model.

\subsection{Problem}\label{problem}

In general, the flaws such as integer overflows and underflows, and flaws brought on by division by zero or modulo zero, are arithmetic flaws. In brief, when an arithmetic expression yields a value that is either greater or smaller than it should be, this is known as an integer overflow or underflow. The typical response in this situation is to silently "wrap around," for example, reducing the value by modulo $2^{32}$ for a 32-bit type. In contrast to Ethereum, where all behavior is well-defined, C/C++ integer operations' out-of-bounds behavior is largely unclear \citep{torres2018osiris}.

\begin{sidewaystable}
\caption{Integer operations' behavior in EVM and Solidity. Both x and y are n-bit mathematical integers, where $x_\infty$, $y_\infty$ stand for the corresponding $\infty$-bit integers \citep{solidity,wood2014ethereum}. $op_s$ stands for signed operations (e.g., +,-,×,÷) and $op_u$ for unsigned operations and also $mod_s$ for singed modulo and $mod_u$ for unsinged modulo.}\label{table1}
\begin{tabular*}{\textwidth}{@{\extracolsep\fill}lcccccc}
\toprule%
\multicolumn{2}{@{}c@{}}{} & \multicolumn{2}{@{}c@{}}{Out of bounds} \\\cmidrule{3-4}%
Integer operation  & In bounds requirement & EVM & Solidity\\
\midrule
$x +_s  y$, $x -_s  y$, $x \times_s  y$ & $x_\infty$ $op$ $y_\infty \in [{-2}^{n-1}$, $2^{n-1}-1]$ & $modulo$ $2^{256}$ & $modulo$ $2^n$\\
$x +_u  y$, $x -_u y$, $x \times_u y$ & $x_\infty$ $op$ $y_\infty \in [0,2^n-1]$ & $modulo$ $2^{256}$ & $modulo$ $2^n$\\
$x \div_s y$ & $y \neq 0 \land (x \neq {-2}^{n-1}  \lor y \neq {-1})$ & $0$ $if$ $y=0$ & $0$\footnotemark[1] $\mid$ $INVALID$\footnotemark[2] $if$ $y=0$\\
 &  & ${-2}^{255}$ $if$ $x={-2}^{255} \land y=-1$ & ${-2}^{n-1}$ $if$ $x={-2}^{n-1} \land y=-1$\\
$x \div_u  y$ & $y \neq 0$ & $0$ & $0$\footnotemark[1] $\mid$ $INVALID$\footnotemark[2]\\
$x$ ${mod}_{s \mid u}$ $y$ & $y \neq 0$ & $0$ & $0$\footnotemark[1] $\mid$ $INVALID$\footnotemark[2]\\
\botrule
\end{tabular*}
\footnotetext[1]{Solidity version $<$ 0.4.0}
\footnotetext[2]{Solidity version $\ge$ 0.4.0}
\end{sidewaystable}

According to Table \ref{table1}, We emit a constraint that is only satisfied if the in-bounds conditions are not met for each arithmetic instruction that might overflow or underflow. Using the addition of two unsigned integers $x$ and $y$ as an example, we can output a constraint to the solver that checks if $x+y > 2^{n-1}$, where $n$ specifies the maximum size of the two values, for example, $64$ in the situation where $x$ is an $uint32$ and $y$ is an $uint64$. Likewise, we examine whether the in-bounds conditions are not met for signed or unsigned division and signed or unsigned modulo. As an illustration, we emit a constraint for signed division that ensures that the divisor cannot be zero. We can determine the likelihood of an arithmetic bug, such as an overflow or a division by zero, by determining if the solver can satisfy any of the emitted constraints under the present route conditions. There are two important findings available. First of all, even though the EVM performs all arithmetic operations modulo $2^{256}$, the following code will quietly wrap around if the value of $x+y$ is greater than $2^{32}-1$.

\begin{minipage}{\hsize}%
\lstset{frame=single,framexleftmargin=-1pt,framexrightmargin=-17pt,framesep=12pt,linewidth=0.98\textwidth}
\begin{lstlisting}
function add(uint32 x, uint32 y) public returns (uint) {
	return x + y;
}
\end{lstlisting}
\end{minipage}

The EVM does not enforce integer operations’ behavior but Solidity does. Second, the zero produces 0 when divided (or modulo) by it. Other programming languages would throw an exception if this happened. However, in previous versions of Solidity to 0.4.0, this produces the zero in Solidity implementations on EVM. Since the majority of developers would anticipate an exception, the Solidity compiler has injected invalid operations since version 0.4.0 to raise an assert-style exception, which forces the EVM to undo all modifications.

\subsection{Motivation}\label{motivation}

The employed neural network models in prior deep learning-based smart contract vulnerability detection tasks are linear, which describes the architecture of the model that has a single structure without any branches and goes through a similar training process. In order to build a single deep learning model with a higher classification accuracy, the number of model layers can be increased. However, doing so will increase the complexity of model, that leads to issues like overfitting and lengthy training times. These issues can be resolved by the development of hybrid networks. By employing two different input representations, where one is the cleaned source code and the other one is the compiled source code as opcode, and two word embedding models to vectorize the inputs and two deep learning models to extract feature, we create our model. Using the benefits of multi-modal hybrid models is the key principle to increase the accuracy of the smart contract vulnerability detection where each hybrid model approach and input representation has its advantages and downsides. It is crucial to use the best input representations and the best hybrid models for detecting the smart contract vulnerability since each deep learning method has unique properties for handling the various tasks, while each input representation has pros and cons. The proposed solution to this issue is to use the benefits of various input representations and hybrid models, so that we propose a new multi-modal hybrid deep learning model entitled as HyMo. For inputs, the model employs two distinct representations and for vectorization, it employs two word embedding techniques. Moreover, for feature extraction, HyMo employs two deep learning models. Our solution's performance is additionally enhanced by different input representations (see Figure \ref{figure1}).

\section{Our Solution}\label{our_solution}

\begin{figure}
\centering
\includegraphics[width=0.9\textwidth]{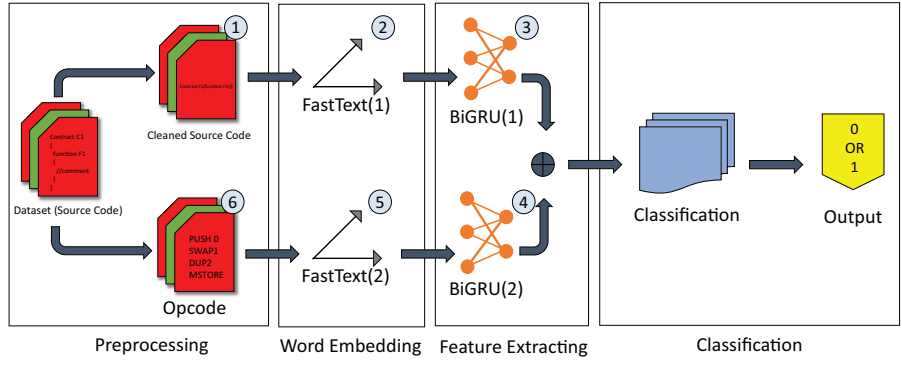}
\caption{HyMo model architecture. In preprocessing phase, we create two inputs for each dataset’s sample of smart contract’s source code, where one input is \#1 cleaned source code and other one is \#6 opcode then we use \#2 FastText word embedding technique to vectorize cleaned source code and \#5 FastText word embedding technique to vectorize opcode and then we feed \#3 BiGRU by FastText values and feed \#4 BiGRU by FastText values then we concatenate these two branches to get final result.}\label{figure1}
\end{figure}

According to Figure \ref{figure1}, four steps of our proposed method are as follows: 

\begin{enumerate}
\item Generating cleaned source code and opcode via preprocessing

\item Using two word embedding technique to map the high-dimensional smart contracts to low-dimensional vectors.

\item Two neural networks to extract the feature values, where they will be concatenated to each other to form super feature vectors.

\item Classification to produce the output where 0 is for immune smart contracts and 1 is for vulnerable ones.
\end{enumerate}

This study concentrates on the first, second, and third phases, combining the benefits of various input data representations, word embedding techniques, and deep learning models to enhance the performance of smart contract vulnerabilities. In part one, we clean the source code of smart contracts and use it as input for a model while using the compiled source code (i.e., opcode) as input for another model. In parts two and three, we discuss the currently well-liked word embedding techniques (e.g., FastText), as well as the currently popular deep learning models (e.g., BiGRU).

\subsection{Preprocessing}\label{preprocessing}

\begin{figure}
\centering
\includegraphics[width=0.5\textwidth]{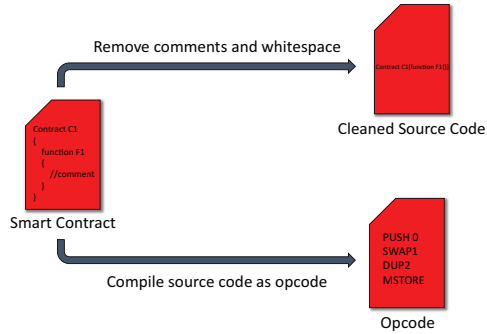}
\caption{Pre-processing structure. Our dataset contains smart contracts’ source code but our inputs are cleaned source code and opcode so we need to prepare dataset’s samples to feed to our model so in preprocessing phase we remove comments and whitespaces from source code and generate cleaned source code for one branch of our model and compile our smart contract sample to opcode for another branch.}\label{figure2}
\end{figure}

In preprocessing phase, we are to produce two different inputs for model where raw input includes many unnecessary parts. Figure \ref{figure2} shows an explicit view from first phase where the produced inputs are as follows:

\begin{enumerate}
\item Clean source code of our model as input (see Figure \ref{figure1}).

\item Compile source code as opcode of our model as input (see Figure \ref{figure1}).
\end{enumerate}

The raw source code for a smart contract will be cleaned by removing all comments and whitespaces, and only the necessary keywords will remain in the source code. In compiling phase, we will look for "pragma solidity x.x.x;" to identify the Solidity compiler version to utilize it for translating the source code into Solidity opcode. Once we have processed the inputs, they feed into word embedding models in next phase.

\subsection{Word Embedding}\label{word_embedding}

The word embedding layer's primary task is to convert the original smart contract into a matrix at the character level so that it conforms to the neural network's input.

By employing the word embedding approach, the tokens are gathered to produce a matrix form that is complied with a deep learning neural network. Both Word2Vec and FastText are processed in the same way in this study.

\subsection{Model Structure}\label{model_structure}

The FastText word embedding model is another word embedding method that is an extension of the Word2Vec model that instead of learning vectors for words directly, it represents each word as an n-gram of characters. FastText word embedding models vectorize the smart contract before both deep learning models begin to feature extraction. In the similar way, both cleaned smart contracts and compiled smart contracts as opcode go through the FastText word embedding models and then BiGRU is used for feature extraction. BiGRU or bidirectional GRU, is a sequence processing model that consists of two GRUs where one taking the input in a forward direction, and the other one in a backward direction. It is a bidirectional recurrent neural network with only the input and forget gates. The unit of BiGRU is set to 300 because the FastText word embedding’s dimension is set to 300-dimensional. The ReLU function performs the activation process and the dropout layer avoids from overfitting. The initial one-way feedback is combined with the two-way feedback provided by BiGRU, which can process input iteratively in both ways. The feature fusion process is carried out through the connection layer following the feature extraction of two branches of the smart contracts. The feature fusion process employs the concatenation method, and the fused feature matrix will perform better than the retrieved feature matrix from a single network. The softmax layer then will be utilized for classification to produce the final outcome. Figure \ref{figure1} elaborates details in schematic form.

\section{Experiments and Results}\label{experiments_and_results}

The employed performance measurements in our study are initially introduced in this section. We employ Adam \citep{kingma2014adam} as an optimizer to update and compute the network parameters where it affects the model training and output in order to approach the optimal values. Adam offers promising approaches for optimizing the resolution of sparse matrix and noise problems, where it gets popular in deep learning applications, particularly for computer vision and natural language processing tasks in recent years. The learning rate of the Adam optimizer is set to 0.001 with reference to the currently well-liked TensorFlow \citep{abadi2016tensorflow} and Keras \citep{keras}. Moreover, dropout layer parameter is set to 0.5 due to the fact that the randomly generated network structure is in its highest, to improve the generalization of the model. Also, the batch size is set to 128 and the epoch is considered 50.

The experiments in this paper are split into two sections: the first section discusses various input representations, deep learning models, and word embedding models to demonstrate the accuracy of our proposed model, and the second section compares the proposed model against the earlier studies to show how well it performs.

\subsection{Dataset}\label{dataset}

\begin{wrapfigure}{l}{5.0cm}
\includegraphics[width=5.0cm]{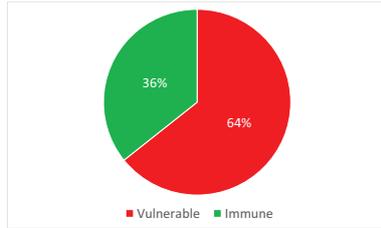}
\caption{Dataset distribution of vulnerable and immune smart contracts.}\label{figure3}
\end{wrapfigure}

There is a labelled real-world Ethereum dataset from smart contracts with vulnerabilities entitled as ScrawlD \citep{yashavant2022scrawld}. The Ethereum community can use ScrawlD to evaluate both new and existing methods for analyzing the vulnerabilities. The 6700 tagged Ethereum smart contracts from the real world are available in ScrawlD. They use Ren et al. \citep{ren2021making} as a methodology to label the smart contracts. We combine the ScrawlD dataset with some labelled smart contracts as arithmetic vulnerabilities from the SmartBugs Dataset-Wild \citep{durieux2020empirical}. Distribution of immune and vulnerable smart contracts are visible in Figure \ref{figure3}.

\subsection{Comparison}\label{comparison}

One of the main contributions of our proposed model is to employ various input representations, word embedding techniques and deep learning models, in order to enhance the performance. As a result, we choose the cleaned source code and the compiled source code (i.e., opcode) as input representations, the currently popular FastText model as word embedding model, and BiGRU as deep learning model.

Under the same procedure and settings, we compare the performance of various input representations, different deep learning models, and various word embedding techniques. We capture data from numerous test sets where deep learning networks modify parameters after each training so that the most accurate model is chosen as our proposed HyMo model. Each model in the experiment underwent just 50 training cycles. The correctness of the test set serves as the performance metric in self-comparison experiment.

\subsubsection{Self-Comparison}\label{self_comparison}

We arrange different input representations, word embedding models and deep learning models to create four single architecture models according to Table \ref{table2}. Our models contain cleaned source code and compiled source code as input representations, FastText and Word2Vec as word embeddings, BiGRU and CNN as deep learning models.

\begin{table}
\caption{Our four single architecture models which every time we use two selected models of them on every hybrid model. $M_i$ stands for one of our 4 designed model to be used in our experiments.}\label{table2}
\begin{tabular*}{\textwidth}{@{\extracolsep\fill}lcccccc}
\toprule%
\multicolumn{1}{@{}c@{}}{Single Model Name} & \multicolumn{3}{@{}c@{}}{Model Architecture} \\\cmidrule{1-1} \cmidrule{2-4}%
& Input & Word embedding & Deep learning model\\
\midrule
M1 & Cleaned source code & Word2Vec & CNN\\
M2 & Cleaned source code & FastText & BiGRU\\
M3 & Compiled source code & Word2Vec & CNN\\
M4 & Compiled source code & FastText & BiGRU\\
\botrule
\end{tabular*}
\end{table}

We choose four hybrid models from all of possible arrange of hybrid models because we want every hybrid model to have both input representations of cleaned source code and compiled source code according to Table \ref{table3}. Names of hybrid models are chosen based on their single architecture models, which contain word embeddings and deep learning models.

\begin{table}
\caption{Each hybrid model contains two single models with different arranges of input representations, word embedding models and deep learning models which the extracted features from every two models will be concatenated together.}\label{table3}
\begin{tabular*}{\textwidth}{@{\extracolsep\fill}lcccccc}
\toprule%
$1^{st}$ and $2^{nd}$ Model & Hybrid Model & Description\\
\midrule
M1, M3 & WCWC (M1\_M3) & Different inputs but same architecture\\
M1, M4 & WCFB (M1\_M4) & Different inputs and architecture\\
M2, M3 & FBWC (M2\_M3) & Different inputs and architecture\\
M2, M4 & FBFB (M2\_M4) & Different inputs but same architecture\\
\botrule
\end{tabular*}
\end{table}

\begin{figure}
\centering
\includegraphics[width=0.9\textwidth]{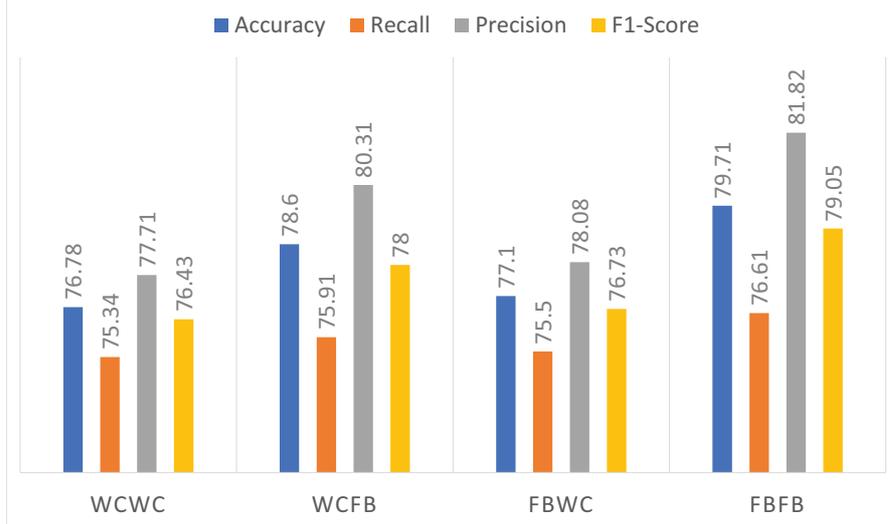}
\caption{Comparison of our proposed models with different arranges of input representations, word embedding models and deep learning models in each model. Evaluation metrics are accuracy, recall, precision and F1-score.}\label{figure4}
\end{figure}

According to Figure \ref{figure4}, our most accurate proposed hybrid model is FBFB (M2\_M4) by accuracy of 79.71\%, recall of 76.61\%, precision of 81.82\% and F1-score of 79.05\%. FBFB hybrid model have better accuracy compared to others because of suitable arrangement of components. In WCWC and WCFB models, we have the same first model which is M1 and different second models which are M3 and M4, while WCFB is more accurate than WCWC so better choice for second model is WCFB which processes the compiled source code as input. Also, in FBWC and FBFB models, we have different first models which are M3 and M4 and same second model which is M2 and while FBFB is more accurate than FBWC so better choice for first model is FBFB which processes the cleaned source code as input. So, we have M2 as best first model and M4 as best second model which both of them used in FBFB and by comparing both best models which are FBFB and WCFB, clearly, we can see that FBFB has both best first model and second model, so FBFB is most accurate model between these four models.

So based on our experiments and achieved results, our proposed model, HyMo is the most accurate with FBFB (M2\_M4) model which contains M2 and M4 single architecture model with the same word embeddings and deep learning model.

\subsubsection{Previous Studies}\label{previous_studies}

\begin{figure}
\centering
\includegraphics[width=0.9\textwidth]{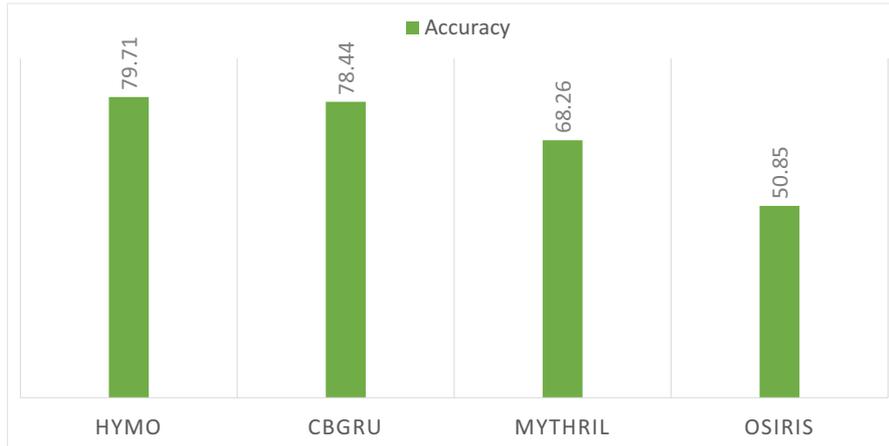}
\caption{HyMo model accuracy in compare with CBGRU model (state-of-the-art of hybrid models) and previous studies such as Mythril and Osiris.}\label{figure5}
\end{figure}

We compared our proposed model with CBGRU \citep{zhang2022cbgru}, a state-of-the-art hybrid model, Mythril \citep{consensys} and Osiris \citep{torres2018osiris}. According to Figure \ref{figure5}, our proposed model HyMo has the best performance by achieving the accuracy of 79.71\% because of using cleaned source code and compiled source code as different input representations and choosing best arrange for our model, FastText as word embedding and BiGRU as deep learning model. CBGRU is at the second place because of using just one representation compared to HyMo, which uses two input representaions. Mythril is at the third place because it relies on concolic analysis, taint analysis and control flow checking of the EVM bytecode \citep{consensys}. Osiris is at the fourth place because it just relies on symbolic execution and taint analysis \citep{torres2018osiris}.

\section{Conclusion and Future Works}\label{conclusion_and_future_works}

In this research, we proposed HyMo as a multi-modal hybrid network model, which is designed based on deep learning. Our experiences on smart contracts arithmetic vulnerabilities, showed that our model has obtained better results than other state-of-the-art hybrid models due to use of multi-modal representations. It does matter to choose best arrange for our model’s components as well as choosing best components like different input representations, word embeddings and deep learning models. The comparison's findings demonstrate that HyMo achieved the considerable performance with an accuracy of 79.71\%. The results demonstrated that HyMo performs better than others in terms of classification accuracy. Moreover, HyMo finds vulnerabilities locally, which is more practical and quicker than using traditional smart contract vulnerability detection methods. In this study, we also empirically showed that the model can extract feature values more accurately when different input representations and word embedding techniques are used. In the subsequent study, we will strive to find all kinds of vulnerabilities in smart contracts and also find the several smart contract vulnerabilities within a single smart contract and enhance the HyMo model's ability to discover flaws in smart contracts with cryptic features.

\section*{Declarations}

\textbf{Funding} Not applicable.\\
\textbf{Competing interests} The authors declare no competing interests.\\
\textbf{Ethics approval} Not applicable.\\
\textbf{Consent to participate} Not applicable.\\
\textbf{Consent for publication} Not applicable.\\
\textbf{Availability of data and materials} The data that has been used during the current study is available in \citep{yashavant2022scrawld} and \citep{durieux2020empirical}.\\
\textbf{Code availability} The code that has been used during the current study is available from the corresponding author on reasonable request.\\
\textbf{Authors' contributions} Mohammad Khodadadi: Data curation, Validation, Investigation, Software, Formal Analysis, Writing - original draft.
Jafar Tahmoresnezhad: Supervision, Conceptualization, Project administration, Resources, Methodology, Data curation, Writing - review and editing.

\bibliography{sn-article}

\end{document}